\documentclass[10pt,letterpaper,twocolumn,prl,showpacs]{revtex4-1}
\usepackage{graphicx,color,textcomp}
\usepackage{mathrsfs,amsmath}   
\usepackage{psfrag,overpic,pst-all}
\usepackage{amssymb}

\begin{document}






\title{Single envelope equation modelling of multi-octave comb arrays in microresonators with quadratic and cubic nonlinearity}

\author{Tobias Hansson$^{1,2}$}
\author{Fran\c{c}ois Leo$^{3}$}
\author{Miro Erkintalo$^{3}$}
\author{Jessienta Anthony$^{3}$}
\author{St\'ephane Coen$^{3}$}
\author{Iolanda Ricciardi$^{4}$}
\author{Maurizio De Rosa$^{4}$}
\author{Stefan Wabnitz$^{1,4}$}
\email{stefan.wabnitz@unibs.it}

\affiliation{$^1$Dipartimento di Ingegneria dell'Informazione, Universit\`a di Brescia, via Branze 38, 25123 Brescia, Italy}
\affiliation{$^2$Department of Applied Physics, Chalmers University of Technology,  SE-41296 G\"oteborg, Sweden}
\affiliation{$^3$The Dodd-Walls Centre for Photonic and Quantum Technologies, Department of Physics, The University of Auckland, Auckland 1142, New Zealand}
\affiliation{$^4$CNR-INO, Istituto Nazionale di Ottica, Via Campi Flegrei 34, 80078 Pozzuoli (NA), Italy}




\begin{abstract}
We numerically study, by means of the single envelope equation, the generation of optical frequency combs ranging from the visible to the mid-infrared spectral regions in resonators with quadratic and cubic nonlinearities. Phase-matched quadratic wave-mixing processes among the comb lines can be activated by low-power continuous wave pumping in the near infrared of a radially poled lithium niobate whispering gallery resonator (WGR). We examine both separate and co-existing intra-cavity doubly resonant second-harmonic generation and parametric oscillation processes, and find that modulation instabilities may lead to the formation of coupled comb arrays extending over multiple octaves. In the temporal domain, the frequency combs may correspond to pulse trains, or isolated pulses.
\end{abstract}



\maketitle


\section{Introduction}

There is currently a great interest in developing microscale, low pump power light sources based on nonlinear parametric frequency conversion for a variety of applications. Early experiments have shown that low-threshold parametric mixing is enabled by the high field confinement of whispering gallery resonators (WGRs) in quadratic quasi-phase-matched (QPM)  \cite{ili03,ili04} or in cubic Kerr  \cite{kippenberg_kerr-nonlinearity_2004} optical microresonators. By using quadratic crystalline microcavities, low-power optical parametric oscillation (OPO) \cite{Savch07}, third \cite{sasa09} and fourth-harmonic generation \cite{moore11} have been demonstrated. Highly tunable OPO can be achieved by using a radially poled WGR \cite{beck11}, or different geometries of domain patterns  \cite{haer10}. It has been proposed that QPM of quadratic frequency conversion in microring resonators may also lead to optical frequency comb generation \cite{wu12}, in analogy with frequency combs from macro cavities based on frequency conversion in a periodically poled lithium niobate (PPLN) nonlinear crystal  \cite{ulvila13,ulvila14,Ricc15a,Ricc15b}.

Nonlinear microresonator based sources of optical frequency combs are a very promising alternative to mode-locked lasers for a variety of potential applications, ranging from high-precision metrology, coherent communications to biomedical and environmental spectroscopy \cite{delhaye_optical_2007,Kippenberg3}. So far,
experimental demonstrations of microresonator-based optical frequency comb sources have been restricted to cubic (or Kerr) type nonlinear materials, which typically requires the use of anomalous chromatic dispersion for achieving broadband frequency combs \cite{delhaye_octave_2011,okawachi_octave-spanning_2011, papp_spectral_2011, herr_universal_2012,bao_nonlinear_2014}. Moreover, even in the presence of the microcavity enhancement of wave mixing, the relatively weak Kerr nonlinearity has so far prevented the demonstration of a low-power microresonator based frequency comb source with a fully integrated pump laser diode. Remarkably, the use of quadratic, as opposed to Kerr, resonators has the potential to lift the anomalous dispersion requirement, and dramatically lower the pump threshold value \cite{Ricc15a,Ricc15b}.

We have recently developed a theoretical description of optical frequency comb generation in both singly resonant \cite{Leo15a}  and doubly resonant \cite{Leo15b} intra-cavity second-harmonic generation (SHG), based on time-domain evolution equations that describe the generation of nonlinearly coupled combs, centered around the fundamental frequency (FF) and the second-harmonic (SH), respectively. In addition to permit the analytical description of modulation instabilities, these models enable computationally efficient numerical simulations of the ensuing  optical frequency combs and corresponding dissipative temporal structures, in particular when cavity-averaged or mean field equations can be derived. 
However, these models are intrinsically limited to the description of combs generated around two distinct carrier frequencies; moreover, they lose validity when the combs start to overlap.

In this work, we present a more general model of \emph{ultra-broadband} frequency comb generation in cavities where several nonlinearities may act simultaneously, based on the single envelope equation (SEE) \cite{GKKD07,conf10,conf10b,victor10,baro12,swab12}. Compared to previous models where a single dominant nonlinear process (e.g., singly or doubly resonant cavity SHG) is involved \cite{Leo15a,Leo15b}, the SEE model has its relative strength in its greater generality, as all nonlinear processes are simultaneously included. 

Here we apply the SEE to study optical frequency comb generation in a doubly resonant, radially poled lithium niobate (RPLN) WGR as demonstrated in Ref.  \cite{beck11}. 
Radial poling enables QPM: by varying the poling period, different quadratic processes can be phase-matched.
We will show by numerical simulations that intra-cavity SHG may lead to modulation instability (MI) and THz rate pulse trains at the fundamental frequency. On the other hand, we predict that, by properly choosing the poling period to phase-match the doubly resonant degenerate OPO process, isolated soliton-like pulses at the fundamental frequency may be produced, corresponding to stable coherent frequency combs at continuous wave (CW) pump power levels in the mW range. The SEE approach is ideally suited to study situations where multiple wave-mixing processes are phase-matched at once. Whenever intra-cavity SHG and OPO coexist, we numerically predict the generation of arrays of nonlinearly coupled combs, extending over multiple octaves, thanks to the interplay of sum-frequency generation (SFG) and difference-frequency generation (DFG) processes. 

\section{Model}
\label{sec:model}

We model the dynamics of optical frequency comb generation by means of an Ikeda-like map \cite{ikeda_multiple-valued_1979} involving the
SEE for describing the
evolution of the broadband envelope \cite{GKKD07,conf10,conf10b,victor10,baro12} $A\:[\sqrt{\textrm{W}}/\textrm{m}]$ of the real electric field $E$ within a waveguide with both quadratic and cubic nonlinearity (i.e., the total nonlinear polarization is $P_{NL}=P_{NL}^{(2)}+P_{NL}^{(3)}=\epsilon_0(\chi^{(2)}E^2+\chi^{(3)}E^3)$, where $\chi^{(2)}$ and $\chi^{(3)}$ are the quadratic and cubic nonlinear susceptibilities and $\epsilon_0$  is the vacuum permittivity). The map is constructed by combining the SEE with boundary conditions that relate the fields between
successive roundtrips and the input pump field
\cite{haelterman_dissipative_1992,coen_modeling_2013}, viz.
\begin{align}
  & \mathscr{F}\left[A^{m+1}(t,0)\right] = \sqrt{\hat{\theta}(\Omega)}\mathscr{F}\left[A_{\textrm{in}}\right] +  \nonumber \\  
  & \sqrt{1-\hat{\theta}(\Omega)}e^{i\phi_0}\mathscr{F}\left[A^m(t,L)\right], \label{eq:Ikeda1}\\
  & \left[\partial_{z} - D(i\partial/\partial t) + \frac{\alpha_d}{2} \right]A^m(t,z)=  \nonumber \\  
   & i\rho_0\left(1+i\tau_{sh}\frac{\partial}{\partial t}\right)p_{NL}(t,z,A^m),
 \label{eq:Ikeda2}
\end{align}
where $p_{NL}$ is the broadband envelope of the nonlinear polarization $P_{NL}$. 
Equation (\ref{eq:Ikeda1}) is written in the Fourier domain so as to enable the modelling of frequency-dependent coupling, with $\hat{\theta}(\Omega)$ describing the
frequency dependent transmission coefficient between the resonator and the bus waveguide, $\Omega=\omega-\omega_0$, $\omega_0$ is a reference frequency (which is set to coincide with the driving pump frequency), and $\mathscr{F}\left[\cdot\right]=\int_{-\infty}^{\infty}\cdot\,e^{i\Omega t}\,\mathrm{d}t$ denotes Fourier transformation. Here the independent variables are the evolution variable $z$, which
is the longitudinal coordinate measured along the waveguide, and $t$ which is the
(ordinary) time. Equation (\ref{eq:Ikeda1}) is the boundary condition that
determines the intra-cavity field $A^{m+1}(t,z=0)$ at the beginning of
roundtrip $m+1$ in terms of the field from the end of the previous
roundtrip $A^m(t,z=L)$ and the pump field $A_{\textrm{in}}$. The path length of
the resonator is assumed to be equal to $L$.
Additionally, $\phi_0 = 2\pi l-\delta_0 \approx (\omega_0-\omega_R)t_R$ is
the linear phase shift, with $\delta_0$ the phase detuning caused by the frequency shift of the pump from the closest cavity
resonance with frequency $\omega_R$ (assumed to
correspond to the longitudinal mode number $l =0$), and $t_R$ is the cavity circulation time.

The SEE (\ref{eq:Ikeda2}) is written in the
reference frame moving at the group velocity at $\omega_0$. Moreover, $\rho_0=\omega_0/(2n_0c\epsilon_0)$, where $n_0 = n(\omega_0)$ is the linear refractive index at $\omega_0$, $\tau_{sh}=1/\omega_0$ is the shock coefficient that describes the frequency dependence of the nonlinearity \cite{GKKD07}, $\alpha_d$ is the distributed linear loss coefficient. The group-velocity dispersion (GVD) operator $D$ reads as
\begin{equation}
 D\left(i\frac{\partial}{\partial t}\right) = \sum_{m\geq2}i\frac{\beta_m}{m!}\left(i\frac{\partial}{\partial t}\right)^{m},
 \label{GVD}
\end{equation}
where $\beta_m = (d^m\beta/d\omega^m)|_{\omega=\omega_0}$ are expansion coefficients of the propagation constant $\beta(\omega)$. In Eq. (\ref{eq:Ikeda2}) the nonlinear polarization $p_{NL}$ is given by the sum of the quadratic and cubic contributions \cite{conf10,conf10b,victor10,baro12}
\begin{align}
& p_{NL}^{(2)}=\frac{\epsilon_0\chi^{(2)}}{2}\left[2|A|^2\exp(i\psi(t,z)) +A^2\exp(-i\psi(t,z))\right],\\
& p_{NL}^{(3)}=\frac{\epsilon_0\chi^{(3)}}{4}\left[3|A|^2A +A^3\exp(-2i\psi(t,z))\right].
\label{p3}
\end{align}
\noindent Here the round-trip index $m$ is implied; moreover, $|A|^2$ only contains nonnegative frequency $\omega_0\geq0$ components \cite{conf10,conf10b}, and $\psi(t,z)=\omega_0t-(\beta_0-\beta_1\omega_0)z$. Note that the SEE approach may be readily extended to include saturable or higher-order nonlinearities in $p_{NL}$.

The quadratic nonlinear coefficient $d\:[\textrm{m}/\sqrt{\textrm{W}}]=\chi^{(2)}/2$, and the nonlinear refractive index $n_2=3\chi^{(3)}/(8n_0)$. In Eq. (\ref{p3}) we have for simplicity neglected the presence of a delayed cubic nonlinear response, or Raman scattering: for a more complete description including Raman terms, see Ref. \cite{baro12}. The real electric field is then obtained as
$E(t,z)=\left(A(t,z)\exp\left\{i\left[\beta_0-\beta_1\omega_0\right]-i\omega_0t\right\}+c.c.\right)/2$, and the total nonlinear
polarization $P_{NL}(t,z)=\left(p_{NL}(t,z)\exp\left\{i\left[\beta_0-\beta_1\omega_0\right]z-i\omega_0t\right\}+c.c.\right)/2$.

In deriving the SEE (\ref{eq:Ikeda2}) from Maxwell equations, besides the scalar plane wave approximation, two additional approximations are necessary: the first is neglecting backward wave propagation; the second is assuming that the linear refractive index $n(\omega)$, which divides the nonlinear polarization in the right-hand-side of  Eq. (\ref{eq:Ikeda2}), remains a constant over the frequency band of interest. Under such assumptions, which are well justified for describing the parametric nonlinear mixing of co-directional waves in transparent (i.e., nonresonant) materials, the SEE (\ref{eq:Ikeda2}) is fully equivalent to Maxwell equations: namely, it does not impose any limitation to the frequency content of a signal \cite{conf10,conf10b,victor10,baro12}. 

As discussed in Appendix A, coupled Eqs. similar to those in Refs.\cite{Leo15a,Leo15b} can be obtained by assuming for Eq. (\ref{eq:Ikeda2}) the ansatz of slowly varying envelopes at the fundamental and the second-harmonic frequencies. Note that, whenever an equal number of modes are considered in the simulation, solving the SEE is as numerically efficient as solving the full coupled cavity map corresponding to models in Refs. \cite{Leo15a,Leo15b}. However, the coupled equation models permit to neglect in the computation all modes except for those contained in a certain bandwidth around the FF and either the SH or the OPO signal and idler, respectively, thus facilitating computational efficiency. The reduction in the number of frequency modes permits a corresponding reduction of computation time. Of course, it is difficult to know a priori how large the comb bandwidth is.

In any case, the main reason why solving the SEE (\ref{eq:Ikeda2}) is computationally demanding when considering a QPM structure is that in this case the integration step should be much smaller than the period of inversion of the sign of the quadratic nonlinearity. For example as we shall see, in the case of intra-cavity SHG the QPM period for RPLN is as short as $\Lambda=18.93\:\mu$m, so that the integration step should be of the order of $1\:\mu$m, whereas the cavity length is typically of the order of $1\:$mm. Corresponding cavity averaged or mean field approximations, which only include the fundamental Fourier component of the QPM structure, permit to use an integration step which is even longer than the cavity length \cite{Leo15a,Leo15b}. This means that a typical difference of more than three orders of magnitude in the integration step, hence the computation time, is involved when comparing mean field models to the more accurate SEE model, where the integration step should be much smaller than the QPM period. A further improvement in efficiency comes from the fact that in the coupled equations approach of Refs. \cite{Leo15a,Leo15b} the modes between the fundamental and the second-harmonic can be neglected: hence, less modes are needed.

\begin{figure}[htbp]
\centering
\includegraphics[width=\linewidth]{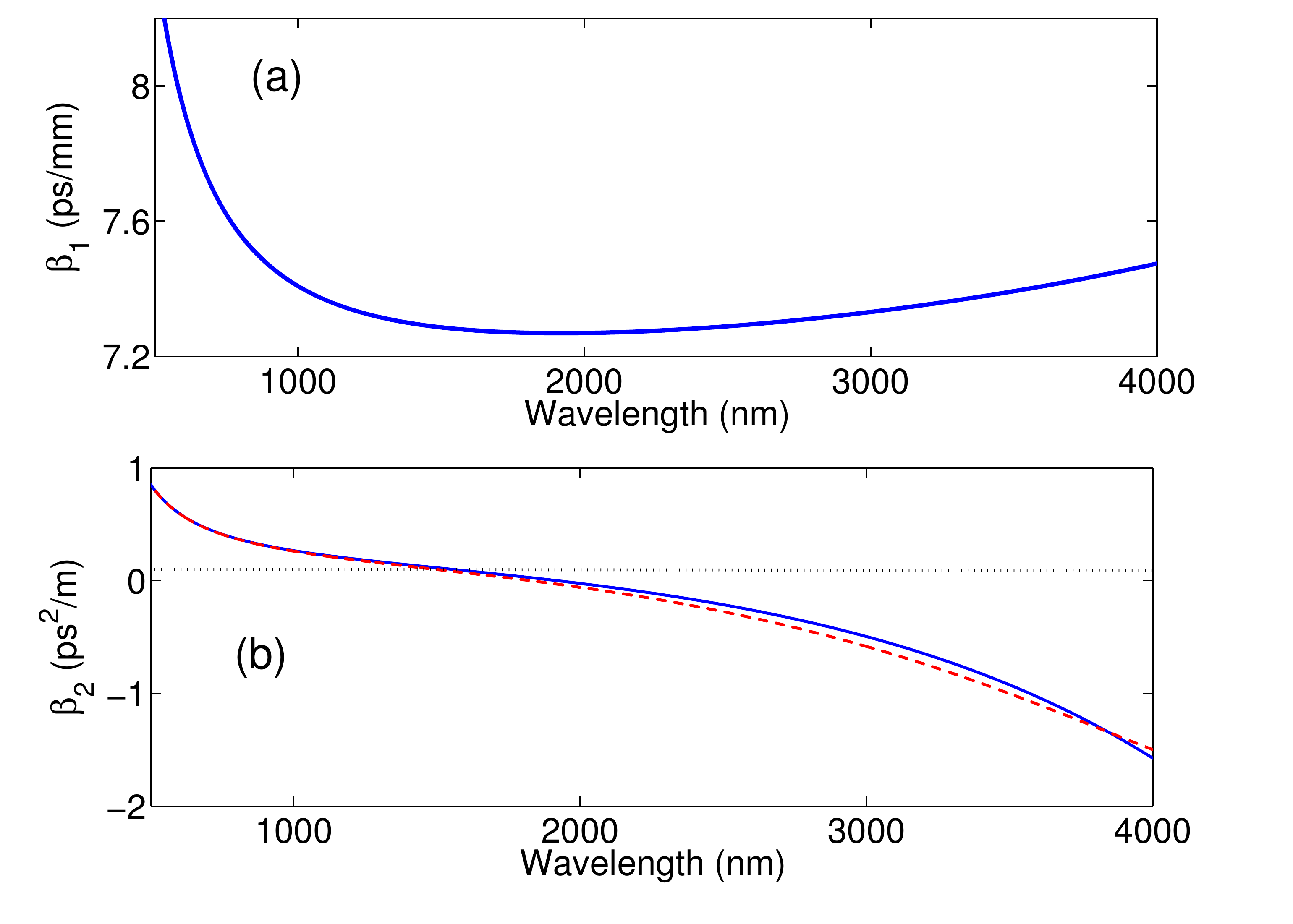}
  \caption{Wavelength dependence of the dispersive properties: (a) group delay $\beta_1$ and (b) dispersion $\beta_2$. Blue solid curves: GVD of lithium niobate WGR: red dashed curve: GVD for TE mode of a RPLN microring with $\mathrm{width}\times\mathrm{height}$ of $12~\mu\mathrm{m}\times7.5~\mu\mathrm{m}$. }
\label{fig:fig1}
\end{figure}

\section{Results}

Equations (\ref{eq:Ikeda1}) and (\ref{eq:Ikeda2}) allow for the full description of parametric frequency comb generation in a RPLN microresonator.
In order to impose QPM among the waves at the center of the respective combs, we included in Eq. (\ref{eq:Ikeda2}) a periodic square wave spatial modulation of the second-order nonlinear coefficient of RPLN $d\equiv  d_{33}=25.2\: \textrm{pm}/\textrm{V}$ \cite{sho,risk03},
with different values of the QPM period $\Lambda$. Moreover, we used the lithium niobate electronic Kerr coefficient $n_2=5.3\times10^{-15}\:\textrm{cm}^2/\textrm{W}$ \cite{leos96}.

In our modeling, we considered a RPLN WGR as in in the experiments of Ref. \cite{beck11}.
We obtained the wavelength dependence of the microresonator GVD by a finite-difference, fully vectorial frequency domain mode solver (Lumerical MODE), using the Sellmeier equation of congruent lithium niobate \cite{jundt97} at the temperature $T=40^\circ$C. The blue solid curves in \ref{fig:fig1}(a) and (b) show the wavelength dependence of the group delay $\beta_1$, and of the GVD $\beta_2$, respectively, for the extraordinary mode of the WGR. The GVD curve virtually coincides with the material dispersion curve.  As it can be seen, the zero dispersion wavelength is at $\lambda_{\textrm{ZDW}}\simeq 1836\:$nm. In Fig.\ref{fig:fig1}(b) we also show as a dashed red curve the estimated dispersion curve of the fundamental TE mode of a RPLN microring \cite{wu12}, with $\mathrm{width}\times\mathrm{height}$ of $12~\mu\mathrm{m}\times7.5~\mu\mathrm{m}$ (so that $A_{\textrm{eff}}\simeq 50\:\mu \textrm{m}^2$) on a lithium tantalate substrate \cite{umeki_2010}. The waveguide dispersion was analytically estimated by the method reported in Ref. \cite{hansson_design_2014}, by assuming a ring radius much larger than the waveguide width. As it can be seen, waveguide dispersion basically leads to a small down-shift of $\lambda_{\textrm{ZDW}}$. 

We numerically simulated the spectral and temporal dynamics of parametric frequency comb generation by solving the SEE (\ref{eq:Ikeda2}) in the frequency domain as a set of coupled ordinary differential equations for the resonator modes, with computationally efficient evaluation of three and four-wave mixing terms in the time domain via the fast Fourier transform method \cite{NCME}. In our simulations, we have used $2^{14}$ cavity modes.

\begin{figure}[htbp]
\centering
\includegraphics[width=\linewidth]{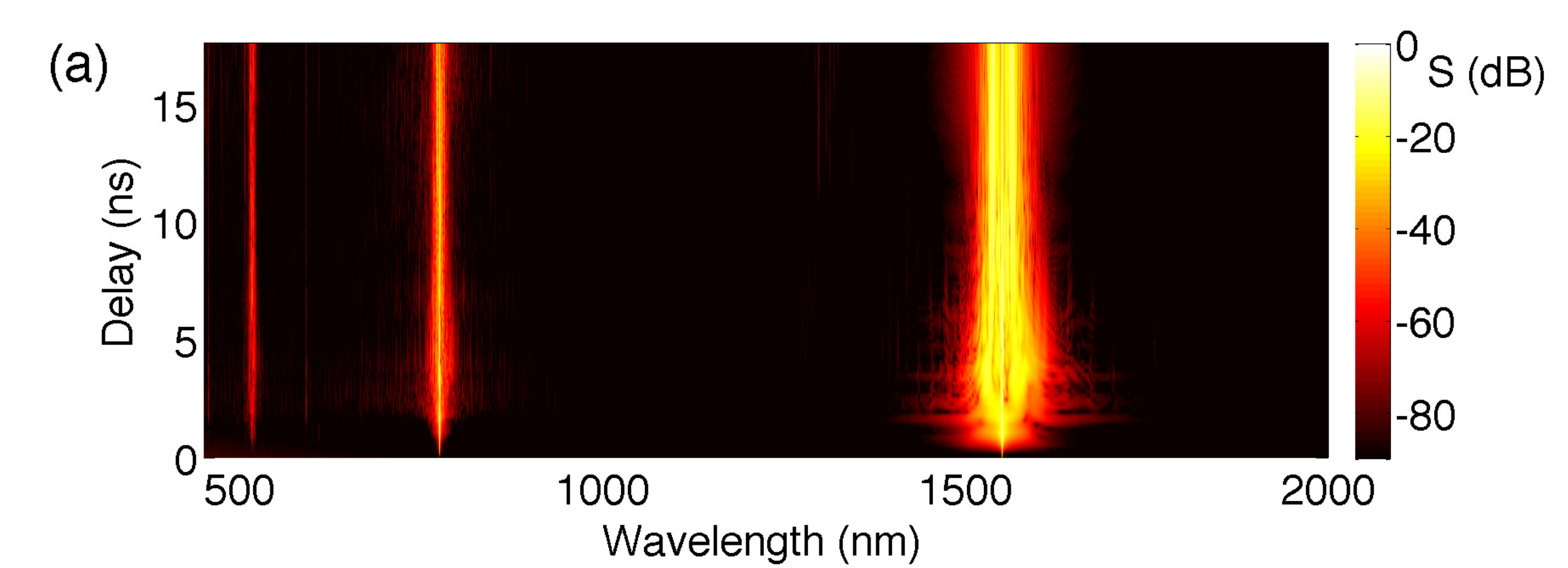}
\includegraphics[width=\linewidth]{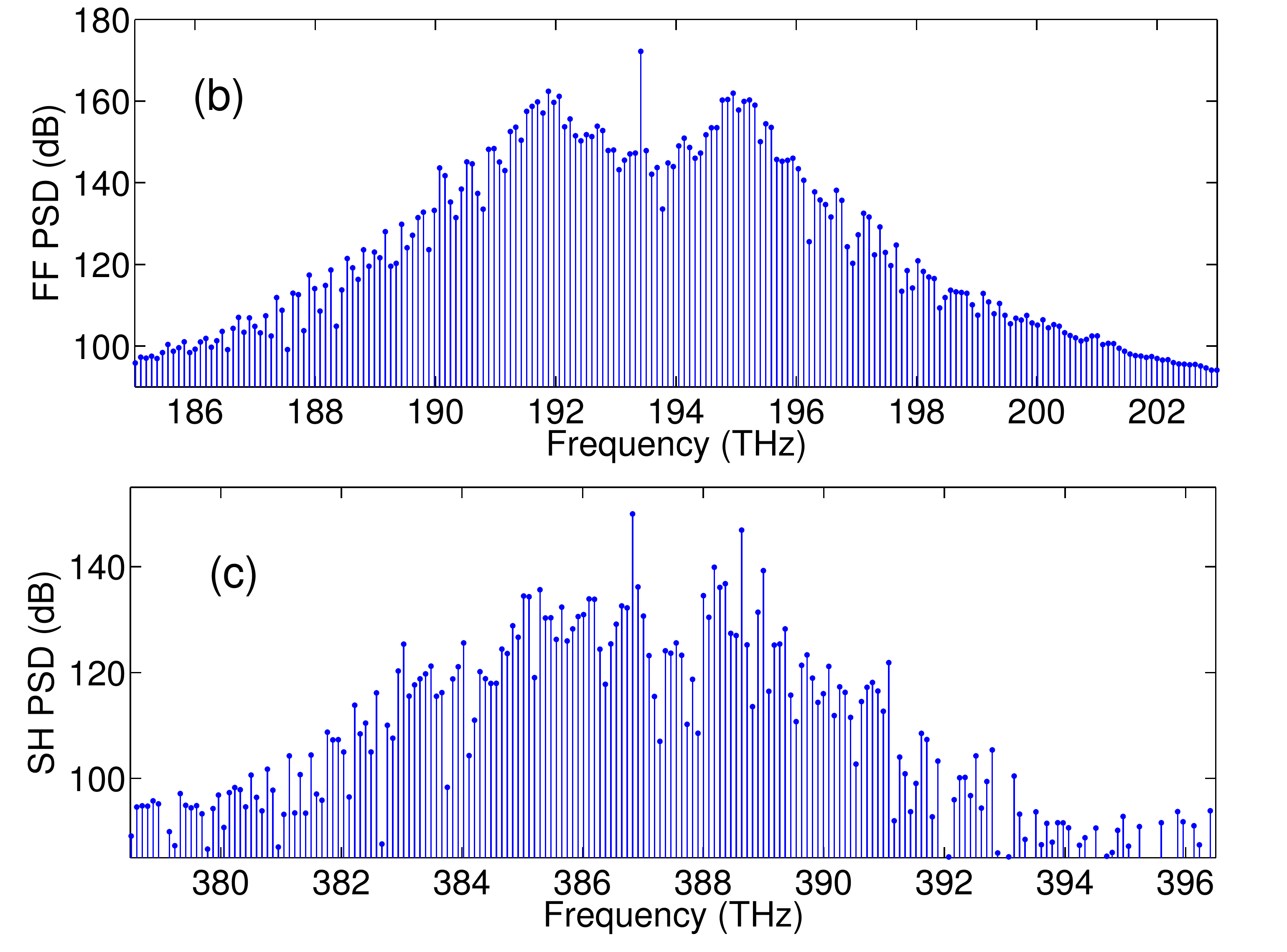}
  \caption{(a) Spectral evolution of intra-cavity intensity showing frequency comb generation at the FF and the SH frequencies; power spectral density (PSD)
 of frequency combs around (b) the FF and (c) the SH frequencies, respectively, after 1600 circulations.}
\label{fig:fig2}
\end{figure}

We considered pumping the RPLN microresonator with $P_{\textrm{in}}=20\:$ mW of input CW fundamental frequency power at $\lambda_0=1550\:$nm, including a
quantum noise (one photon per mode) background. The microresonator, with quality factor $Q=\pi n_gL/(\alpha'\lambda_0)\simeq 10^{6}$, where $\alpha'=(\alpha_dL+\theta)/2$, was operated in the critical-coupling regime, i.e., we set $\theta=\alpha_d L$, so that $\theta=0.0067$ at the pump wavelength $\lambda_0=1550\:$nm. 
Note that, for simplicity, we considered a frequency independent coupling coefficient $\hat{\theta}(\Omega)\equiv \theta$.
We took into account the strong material absorption of lithium niobate in the blue and in the mid-infrared (MIR) sides of the spectrum \cite{myers96}, respectively, by assuming 
a frequency dependent loss of the form $\hat{\alpha}(f=(\omega_0+\Omega)/(2\pi))\equiv \mathscr{F}\left[\alpha_d \right] =\alpha_{1550}\left(1+\exp\left\{\left[f-c/400\right]/b_1\right\}+
\exp\left\{-\left[f-c/4000\right]/b_2\right\}\right)$, where $b_1=30$ THz and $b_2=4$ THz, and  $\alpha_{1550}$ is the linear attenuation of the lithium niobate waveguide at 1550 nm. 

\begin{figure}[htbp]
\centering
\includegraphics[width=\linewidth]{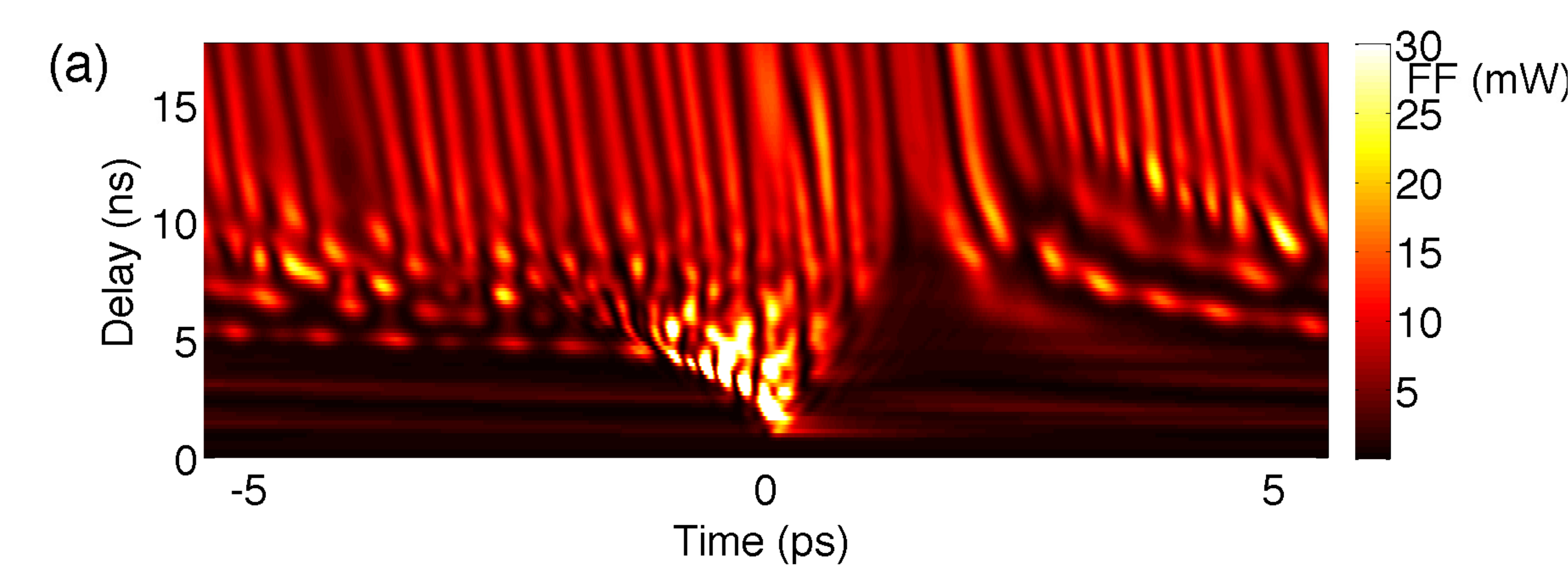}
\includegraphics[width=\linewidth]{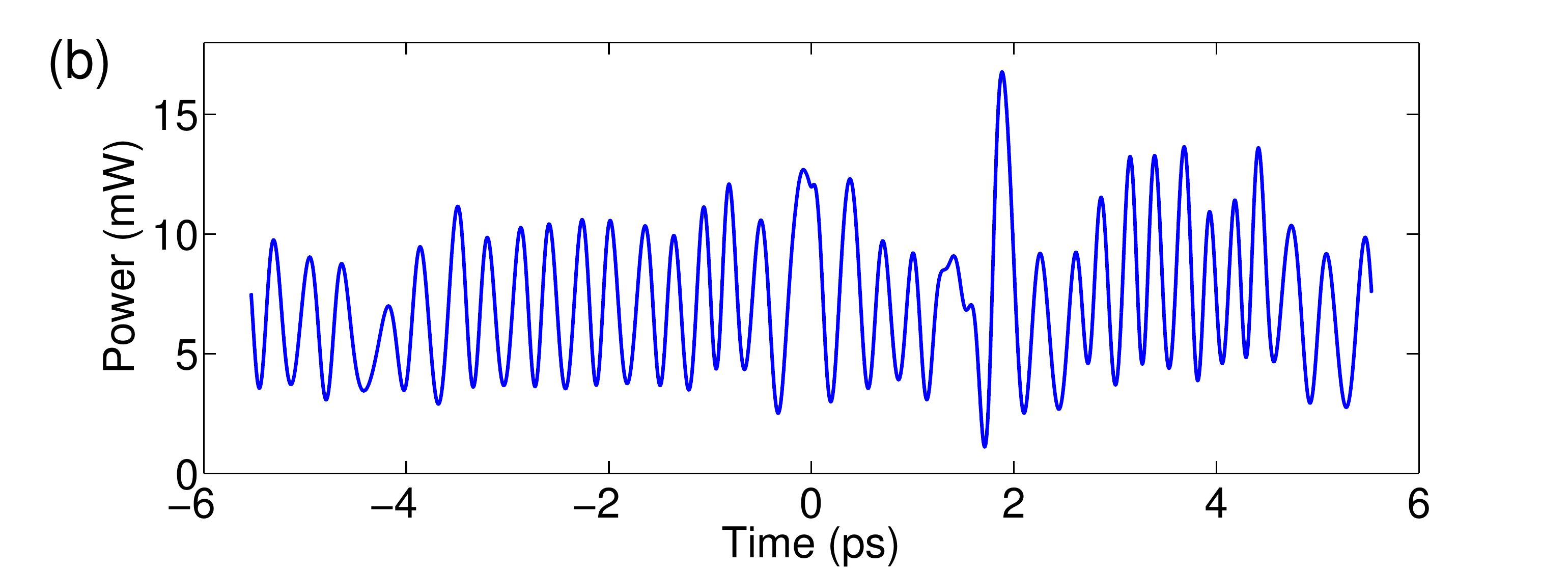}
  \caption{(a) Temporal evolution of power at the FF frequency in the presence of intra-cavity SHG;
 (b) temporal power profile at the FF frequency after 1600 circulations.}
\label{fig:fig3}
\end{figure}

\subsection{Intra-cavity second harmonic generation}

Let us discuss first the dynamics of simultaneous frequency comb generation at the FF and the SH, obtained for a QPM period $\Lambda_{\textrm{SH}}=18.93\:\mu$m, which ensures
quasi-phase-matching of intra-cavity SHG or $\beta_0^{\textrm{SH}}-2\beta_0^{\textrm{FF}}=2\pi/\Lambda_{\textrm{SH}}$. The WGR circumference is equal to $L\approx 1.5 $ mm, corresponding to the radius $R\simeq 0.24 $ mm. Note that the hypothesis of a frequency independent coupling coefficient $\theta$ means that the cavity SHG can be made doubly resonant by properly choosing the WGR diameter (or by tuning its temperature), to have simultaneous resonance at the FF and the SH. In order to allow for the possibility of generating a coherent octave spanning comb comprising both the FF and the SH, we finely adjusted the cavity length $L$ so that the SH frequency $f_{\textrm{SH}}=386.83\:$THz is separated from the FF frequency
$f_{\textrm{FF}}=193.41\:$THz by an integer number \textit{N} of FSRs, namely,
 \textit{N} =2140 FSR and the FSR=90.38 GHz (at the FF wavelength of 1550 nm). In the simulations we considered a time window of 11.064 ps, which corresponds to a frequency grid of a single FSR in the spectral domain, so that both the FF and the SH belong to the same numerical grid. Quite interestingly, our simulations reveal that coupled frequency combs are also generated, with nearly the same efficiency, whenever the SH frequency does not belong to the grid. In fact, quadratic frequency combs are generated via SFG and OPO processes involving many frequency components around the FF and the SH \cite{Ricc15a,Ricc15b}. 

 %
\begin{figure}[b]
\centering
\includegraphics[width=\linewidth]{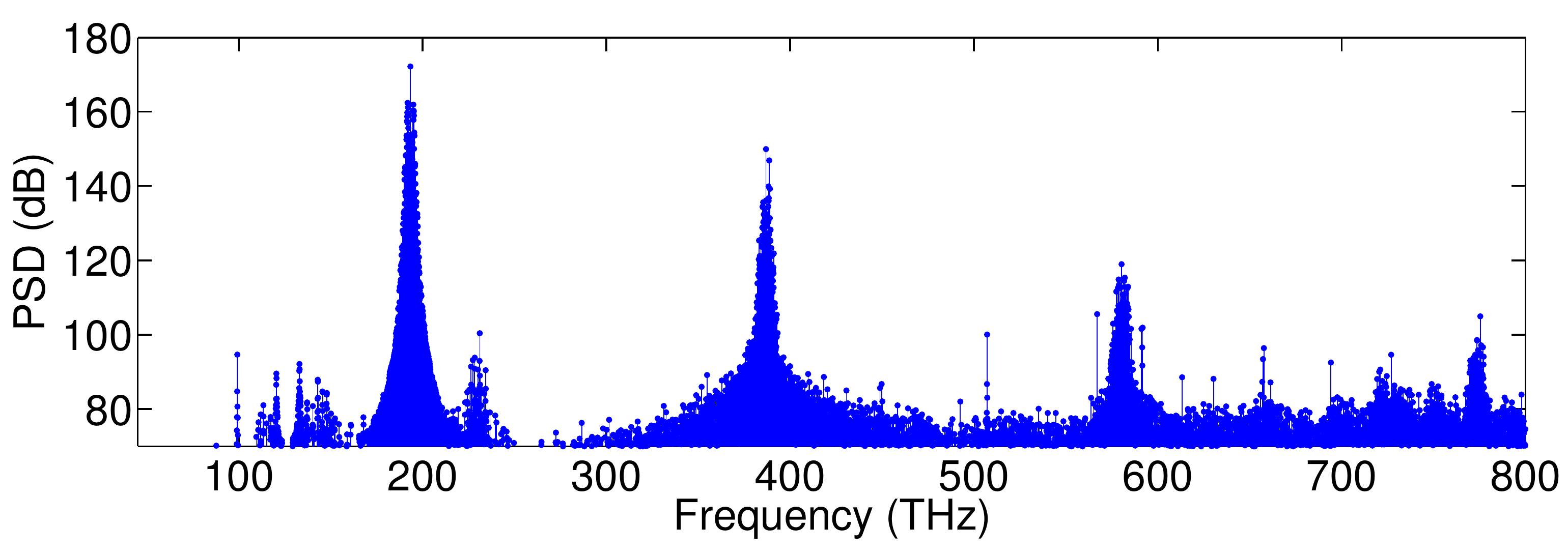}
  \caption{Broadband intra-cavity SHG spectrum as in Fig. \ref{fig:fig2}, after 1600 circulations, shown over a broader spectral window.}
\label{fig:fig3tris}
\end{figure}

Figure \ref{fig:fig2}(a) shows the spectral evolution of the intra-cavity intensity with the FF cavity detuning $\delta_0=0.0081$, which corresponds to the dimensionless detuning $\Delta=\delta_0/\theta=1.2$. The considered value of the detuning is intermediate within the range considered in the coupled equations simulations in Ref. \cite{Leo15b}. We also considered different values of the detuning (for example zero detuning) and obtained qualitatively similar results. In Fig. \ref{fig:fig2}(a), the vertical axis indicates the propagation delay $\beta_1z$ that increases with the number of cavity circulations. It can be clearly seen that frequency combs are generated around both the FF and the SH frequencies  \cite{Ricc15a,Ricc15b}. The corresponding intra-cavity power spectral densities of frequency combs around the FF and SH frequencies after 1600 field re-circulations are also shown in Fig. \ref{fig:fig2}(b) and Fig. \ref{fig:fig2}(c), respectively.

Note that SHG occurs in the RPLN microresonator in the regime of large group-velocity mismatch (GVM). In fact, 
$\beta_1^{\textrm{FF}}=7.28 \:$ps/mm, $\beta_1^{\textrm{SH}}=7.59 \:$ps/mm and $\beta_2^{\textrm{FF}}=0.1 \:\textrm{ps}^2/\textrm{m}$, and the dimensionless
GVM parameter \cite{Leo15a,Leo15b} $g_{\textrm{SH}}\equiv (\beta_1^{\textrm{SH}}-\beta_1^{\textrm{FF}})\sqrt{2L/(\alpha |\beta_2^{\textrm{FF}}|)} \simeq 642$, which corresponds to the regime of walk-off activated MI for the FF. In fact the frequency comb spectrum in Fig. \ref{fig:fig2}(b) shows the presence of two MI peaks symmetrically positioned around the FF with a frequency detuning of about $\pm1.5\:$THz from the pump, in qualitative agreement with analytical predictions \cite{Leo15a,Leo15b}.

 %
\begin{figure}[t]
\centering
\includegraphics[width=\linewidth]{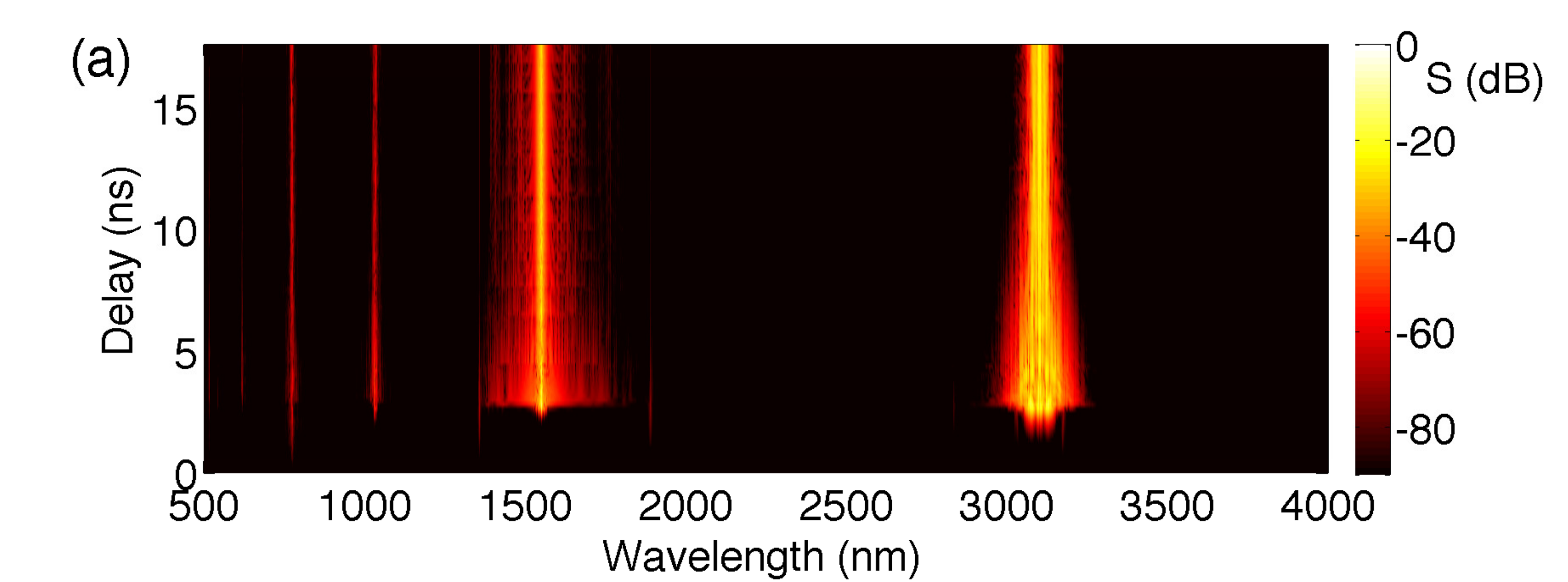}
\includegraphics[width=\linewidth]{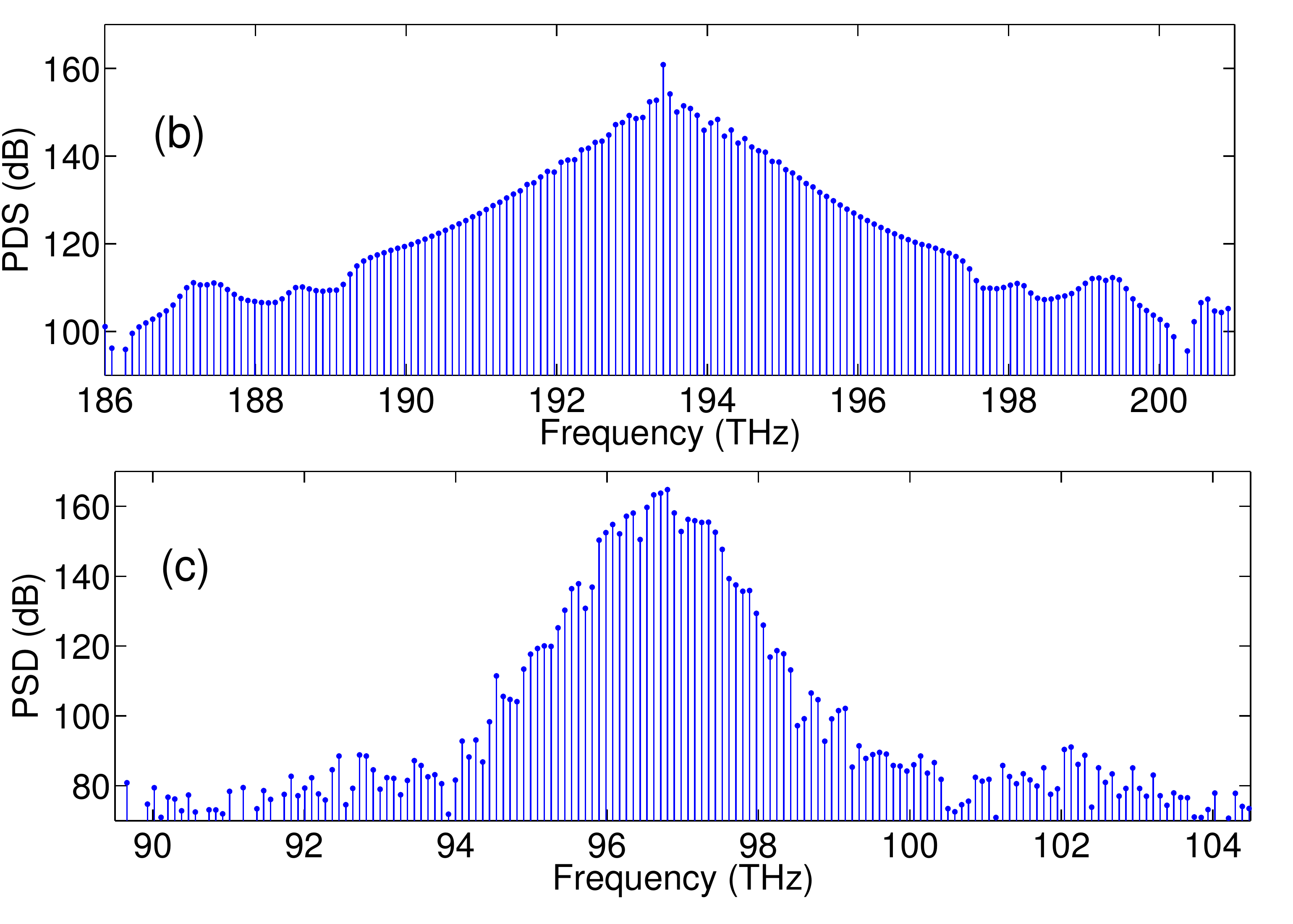}
\includegraphics[width=\linewidth]{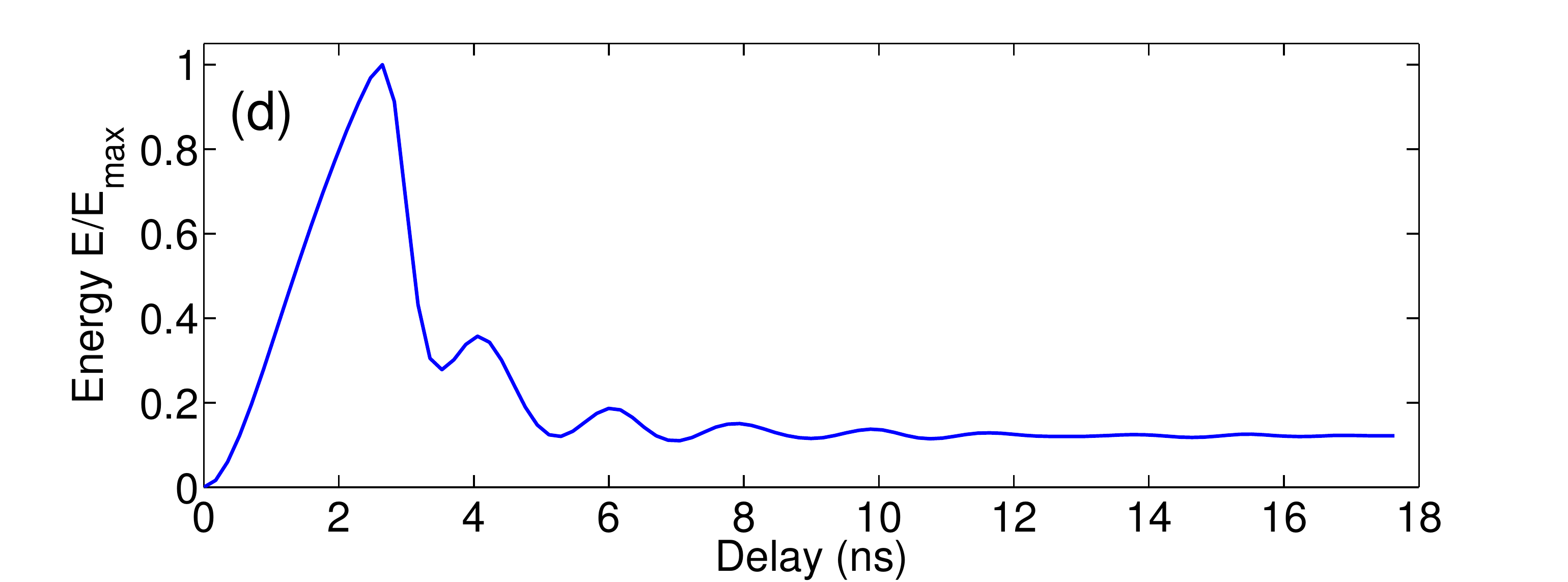}
  \caption{(a) Spectral evolution of intra-cavity power showing frequency comb generation at the FF and the OPO frequencies;
 comb power spectral density around the (b) FF and the (c) OPO frequencies after 1600 circulations; (d): evolution of dimensionless intra-cavity energy. }
\label{fig:fig4}
\end{figure}

The time domain evolution of the intra-cavity power of the frequency comb generated at the FF is illustrated in Fig. \ref{fig:fig3}(a). Here we plot $|A|^2A_{\textrm{eff}}\theta$ (where the field $A$ has been filtered by a square filter around the FF comb), in order to visualize the power level from the microresonator when the intra-cavity field is extracted from a drop port. 
As can be seen, the initial stage of comb formation leads to a narrow intense pulse at the center of the temporal window, that subsequently breaks up into a nearly periodic wavetrain. The snapshot of the temporal power profile at the FF after 1600 ring circulations in Fig. \ref{fig:fig3}(b) confirms the presence of a nearly sinusoidal oscillation with a slowly varying envelope on a background, with the main period of about $670\:$fs.

Although, as shown in Fig. \ref{fig:fig2}(a), for an input CW pump power of $P_{\textrm{in}}=20\:$mW the intra-cavity SHG process leads to the generation of two main combs around the FF and the SH, the broadband spectrum of light circulating in the cavity reveals the presence of an additional comb at the third-harmonic (TH) frequency, as displayed in Fig. \ref{fig:fig3tris}. 

In order to understand the possible role of the cubic nonlinearity (or Kerr effect) in the generation of optical frequency combs,
we have solved Eqs. (\ref{eq:Ikeda1})-(\ref{eq:Ikeda2})
with a quadratic nonlinearity only, that is, we have set $\chi^{(3)}=0$. 
Our simulations show that that cubic nonlinearity has virtually no influence on combs induced by intra-cavity SHG. In particular, the comb generated at the TH is essentially unaffected by the third-harmonic contribution of the cubic nonlinear polarization. This means that the TH comb is generated by the SFG of the FF and the SH.

\subsection{Optical parametric oscillation}

Next, we consider frequency comb generation from the FF to the MIR, assuming a poling period $\Lambda_{\textrm{OPO}}=34.4$~$\mu$m for achieving QPM for the degenerate OPO at $\lambda_s =3100$~nm, or $\beta_0^{\textrm{FF}}-2\beta_0^{\textrm{OPO}}=2\pi/\Lambda_{\textrm{OPO}}$. 
Since the OPO signal is well within the transparency window of lithium niobate, again the cavity can be considered as doubly resonant (for both the FF and the OPO). We keep here the $\textrm{FSR}=90.38\:$GHz, and set to zero the FF cavity detuning $\delta_0=0$. Note that the GVM between FF and OPO signal reads $g_{\textrm{OPO}}\equiv(\beta_1^{\textrm{OPO}}-\beta_1^{\textrm{FF}})\sqrt{2L/(\alpha |\beta_2^{\textrm{FF}}|)} \simeq 127$.

\begin{figure}[htbp]
\centering
\includegraphics[width=\linewidth]{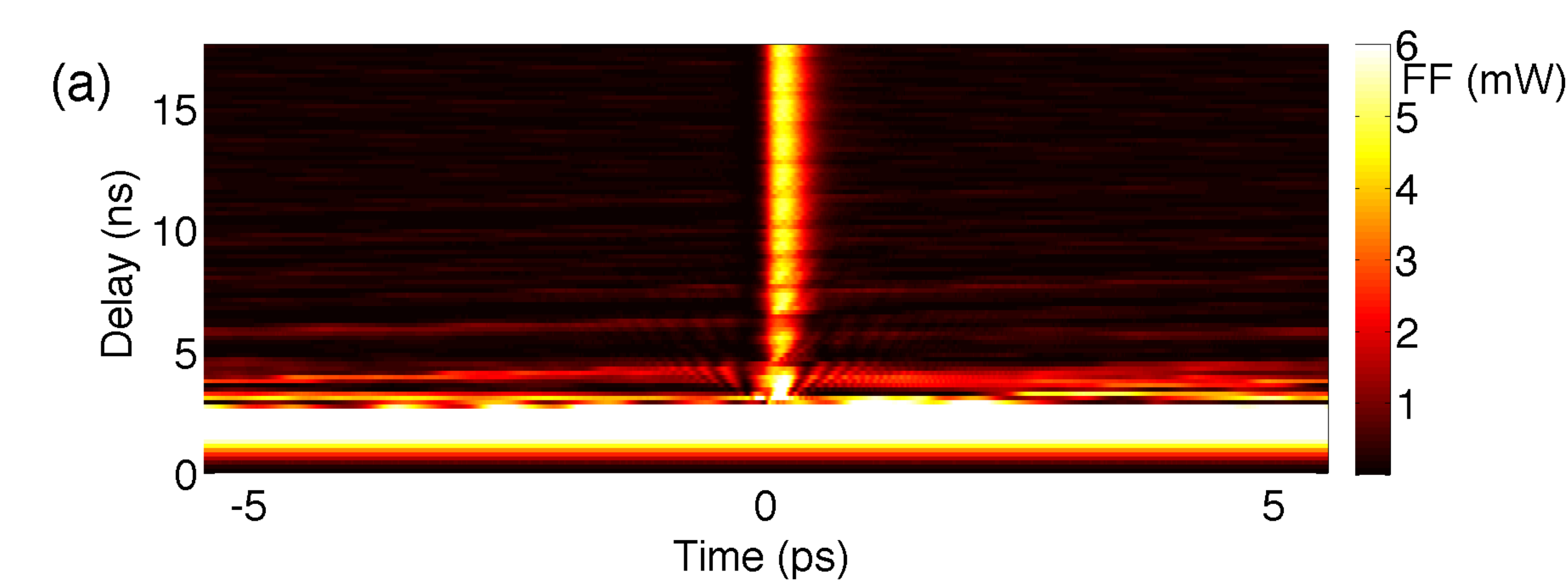}
\includegraphics[width=\linewidth]{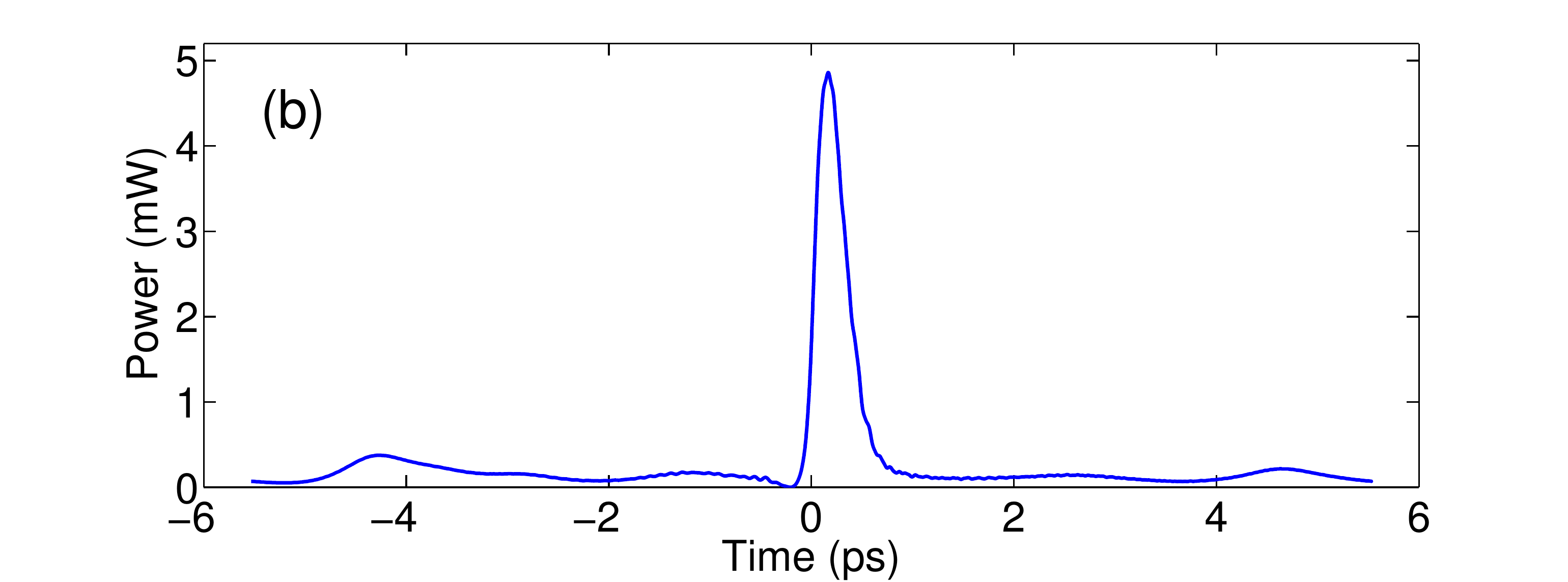}
  \caption{(a) Temporal evolution of the power at the FF frequency in the presence of the OPO comb;
  (b) temporal power profile at the FF frequency after 1600 circulations.}
  \label{fig:fig5}
\end{figure}

The plot in Fig. \ref{fig:fig4}(a) shows that, for the pump power of $P_{\textrm{in}}=20\:$mW the intra-cavity optical parametric generation process generates two main combs, one around the FF frequency $f_0=193$ THz [see Fig. \ref{fig:fig4}(b)], and one around the OPO frequency $f_0/2$ [see Fig. \ref{fig:fig4}(c)]. Moreover, additional low-intensity frequency combs are also generated around the SH frequency $2f_0$ and around $3f_0/2=289$ THz, arising from SFG between the OPO and the FF frequencies. The triangular shape of the FF spectral profile at 1600 round-trips in Fig. \ref{fig:fig4}(b) suggests that a coherent (i.e., phase locked) hyperbolic secant type of pulse train is generated. The plot of Fig. \ref{fig:fig4}(d) shows that the intra-cavity energy of the field initially shoots (after about 2.5 ns) to a peak value, and later stabilizes to a fixed value after about 1000 round-trips (or 10 ns). 

\begin{figure}[htbp]
\centering
\includegraphics[width=\linewidth]{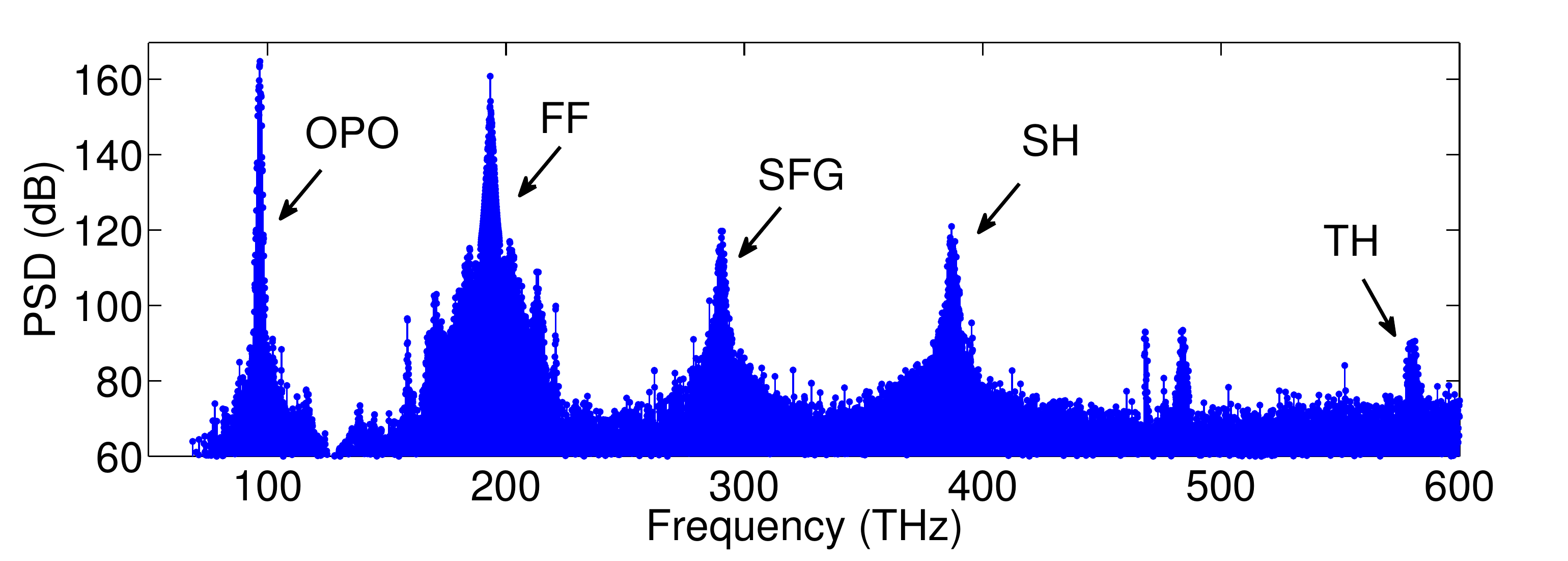}
  \caption{Broadband intra-cavity OPO spectrum with CW pump power $P_{\textrm{in}}=20\:$mW at 1600 round-trips.}
\label{fig:fig5bis}
\end{figure}

Next in Fig. \ref{fig:fig5} we display the evolution in the time-domain of $|A|^2A_{\textrm{eff}}\theta$ (where again the field $A$ has been filtered by a square filter around the FF comb), associated with the FF comb over 1600 round-trips, which indeed confirms that the OPO process leads to the generation of a single stable intra-cavity solitary pulse per round-trip with about 0.3 ps duration, sitting on a relatively low power background. The peak intra-cavity energy point observed in Fig. \ref{fig:fig4} after 2.5 ns, corresponds in Fig. \ref{fig:fig5} to the generation of an unstable intra-cavity CW field (with about 12 mW of out-coupled power). Subsequently, the CW background switches to a low value, and an individual stable pulse with a peak power slightly below 5 mW is formed. The numerically observed generation of a solitary pulse in the doubly resonant, CW pumped degenerate OPO process is closely related with the analytical prediction by Longhi, who obtained an hyperbolic secant solitary wave solution of the singly resonant degenerate OPO process \cite{longhi_1995}.

A clearer view of the broadband intra-cavity spectrum after 1600 cavity circulations is provided by Fig. \ref{fig:fig5bis}, which shows well the secondary comb resulting from SFG of the OPO signal and the FF. In addition, harmonic (SH, TH) secondary combs are also visible in Fig. \ref{fig:fig5bis}.

\subsection{Multiple combs}

\begin{figure}[t]
\centering
\includegraphics[width=\linewidth]{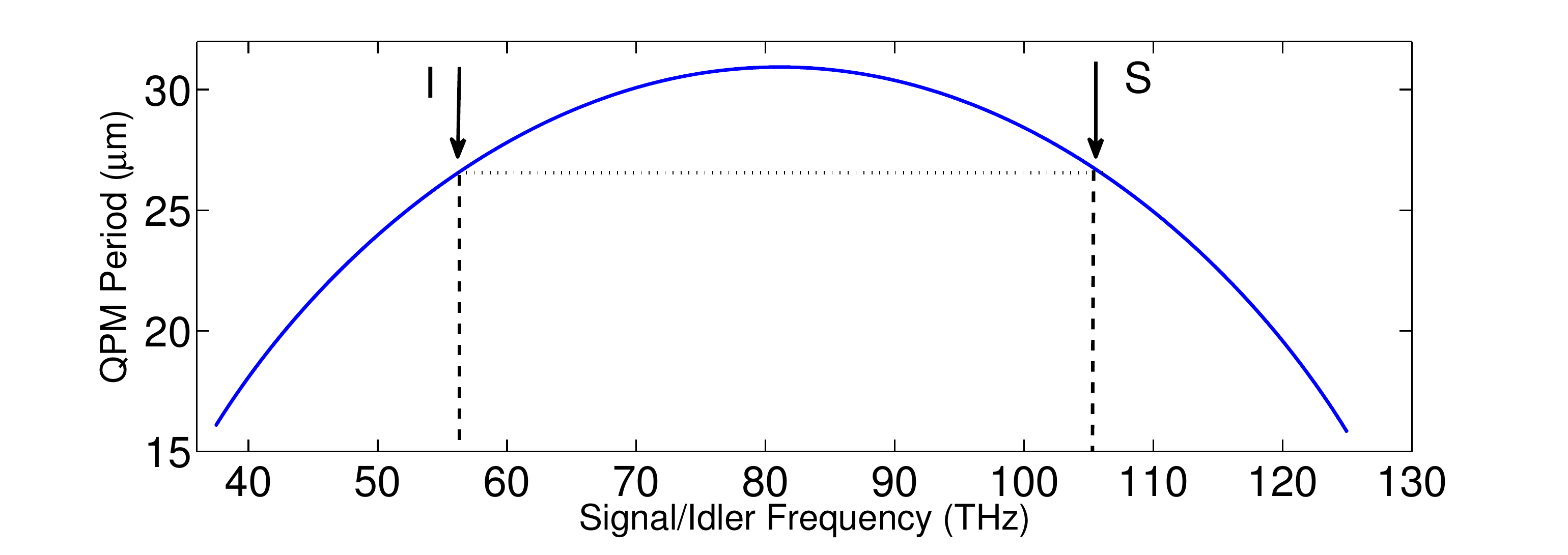}
  \caption{Dependence of OPO signal and idler frequency on QPM period with CW pump at 1850 nm. Dashed line corresponds to the QPM period for SHG.}
\label{fig:fig6}
\end{figure}

In previous sections we have separately considered the processes of intra-cavity SHG and OPO as comb generation mechanisms. The power of the SEE approach is however better exploited when multiple wave mixing processes are simultaneously phase matched, and thus concur to the generation of an ultra-broadband array of multiple frequency combs. For example, with a FF at $\lambda_0=1550\:$nm, it turns out that the QPM period $\Lambda_{\textrm{SH}}=18.93\:\mu$m also leads to phase matching of the non-degenerate OPO process involving the idler and signal with frequencies $f_{\textrm{I}}=41\:$THz (or 7310 nm) and $f_{\textrm{S}}=f_{\textrm{FF}}-f_{\textrm{I}}\simeq 152\:$THz (or 1970 nm). As the idler falls well outside the transparency window of lithium niobate, to investigate the generation of multiple parametric combs we moved the CW pump wavelength to $\lambda_0=1850\:$nm, or $f_{\textrm{FF}}=162\:$THz. Correspondingly, the QPM period of intra-cavity SHG is $\Lambda_{\textrm{SH}}=25.56\:\mu$m. In this case, as illustrated in Fig. \ref{fig:fig6}, the SHG poling period also leads to QPM of the non-degenerate OPO process, involving the generation of an idler and a signal with frequencies $f_{\textrm{I}}=41\:$THz (or 5353 nm) and $f_{\textrm{S}}=f_{\textrm{FF}}-f_{\textrm{I}}\simeq106\:$THz (or 2828 nm), respectively,
so that $\beta_0^{\textrm{FF}}-\beta_0^{\textrm{S}}-\beta_0^{\textrm{I}}=2\pi/\Lambda$.  

\begin{figure}[b]
\centering
\includegraphics[width=\linewidth]{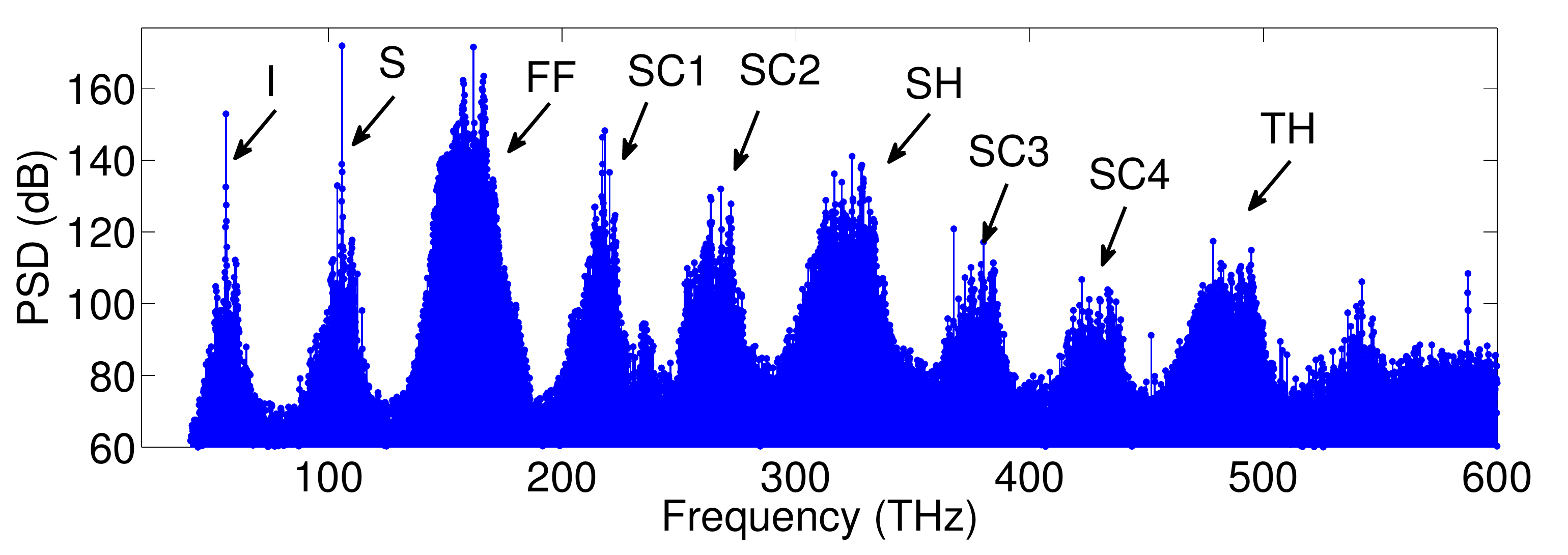}
  \caption{Broadband intra-cavity spectrum with CW pump at 1850 nm and power $P_{\textrm{in}}=100\:$mW at 1600 round-trips.}
\label{fig:fig7}
\end{figure}

In this case, the cavity $\textrm{FSR}=91.71\:$GHz. In order to overcome the power threshold for the non-degenerate OPO process, we increased the CW pump power to $P_{\textrm{in}}=100\:$ mW. The resulting broadband intra-cavity power spectral density after 1600 circulations is shown in Fig. \ref{fig:fig7}: as can be seen, in addition to the FF, SH and TH combs, two additional combs are also generated around the signal and idler frequencies. Note that the intra-cavity signal power is nearly equal to the residual power at the pump wavelength, in spite of the nearly two orders of magnitude larger linear attenuation at the idler frequency (i.e., $\hat{\alpha}(f_{\textrm{I}})/\alpha_{1550}\simeq 110$).

Fig. \ref{fig:fig7} also show the presence of several secondary combs between the FF and the SH, and between the SH and the TH, respectively. These combs are generated by SFG and DFG processes, and demonstrate a novel possibility of generating a multi-comb array extending from the MIR into the blue with a single FSR. For example, SFG between the FF (SH) and the idler leads to secondary comb SC1 (SC3) centered at $f_{\textrm{SC1}}=f_{\textrm{FF}}+f_{\textrm{I}} \simeq 218\:$THz ($f_{\textrm{SC3}}=f_{\textrm{SH}}+f_{\textrm{I}} \simeq 380\:$THz). While DFG between the SH (TH) and the idler leads to the secondary comb SC2 (SC4) centered at $f_{\textrm{SC2}}=f_{\textrm{SH}}-f_{\textrm{I}} \simeq 268\:$THz ($f_{\textrm{SC4}}=f_{\textrm{TH}}-f_{\textrm{I}}\simeq 430\:$THz).  A detailed investigation of the characteristics and dynamics of the multiple comb field is left for future study.
%
%



%
%


\section{Conclusions}

In summary, we have applied the single envelope equation approach to numerically simulate the process of optical frequency comb generation in a RPLN WGR pumped by a relatively low power near-infrared CW beam. By adjusting the QPM period in order to phase-match either the SH or the OPO process, we predicted the generation of coherent pulse trains or solitary pulse frequency combs around the fundamental frequency, coupled with a parametrically generated comb in either the visible or the MIR region.

The SEE approach also permits to reveal the generation of secondary combs, that interact with the primary combs via the interplay of SFG and DFG processes, so that an array of multiple combs is generated over a spectral region covering several octaves. These results point to the possibility of using quadratic nonlinear microresonators for flexible and low-power frequency comb generation over an ultra-wide frequency range, which is of high potential interest for multiple scientific and technological applications.

\section{Funding Information}

Italian Ministry of University and Research (MIUR) (2012BFNWZ2, and Progetto Premiale QUANTOM---Quantum
Opto-Mechanics); Swedish Research Council (SRC) (2013-7508); Rutherford Discovery Fellowships of the Royal Society of New Zealand; Marsden Fund of The Royal Society of New Zealand; Finnish Cultural Foundation.

\appendix
\section{Appendix}

In this appendix we consider the relation of the SEE model to the doubly resonant cavity SHG map for the fundamental and second harmonic field presented in Ref.~\cite{Leo15b}. For simplicity we neglect $\chi^{(3)}$ terms and limit the derivation to the quadratic nonlinearity only. 

We begin by assuming that the field envelope is dominated by components at the FF and SH frequencies, $A_1$, and $A_2$, respectively, so that the total field may be expanded as
\begin{equation}
  A^m(t,z) = A_1(t,z) + A_2(t,z)\exp{[i(Kz-\omega_0 t)]}
  \label{eq:ansatz}
\end{equation}
where $K$ is a constant to be determined for consistency with the coupled equations model of Ref.~\cite{Leo15b}, and proceed to insert this ansatz into the quadratic polarization term. Considering only FF and SH frequency components we find that the polarization produces two relevant nonlinear terms, viz.
\begin{align}
  & 2|A|^2e^{i\Psi(t,z)} \sim 2A_2A_1^*e^{iKz - i(\beta_0-\beta_1\omega_0)z}, \\ \nonumber
  & A^2e^{-i\Psi(t,z)} \sim A_1^2e^{-i\omega_0 t+(\beta_0-\beta_1\omega_0)z}. \nonumber
\end{align}
Inserting these into Eq. (\ref{eq:Ikeda2}) and projecting onto each frequency component we obtain the two equations
\begin{equation}
  \left[\partial_z - D(\omega) + \frac{\alpha_d}{2}\right]A_1 = i\kappa A_2A_1^*e^{iKz - i(\beta_0-\beta_1\omega_0)z}
\end{equation}
\begin{equation}
  \left[\partial_z +iK - D(\omega+\omega_0) + \frac{\alpha_d}{2}\right]A_2 = i\kappa A_1^2e^{-iKz+(\beta_0-\beta_1\omega_0)z},
\end{equation}
where $\kappa = \omega_0\chi^{(2)}/(2n_0c)$ and we have used that $i\tau_{sh}\partial_t A_1^2(t,z)e^{-i\omega_0t} \approx A_1^2(t,z)e^{-i\omega_0t}$ when evaluating the self-steepening term. The group-velocity dispersion operator $D$ (see Eq. (\ref{GVD})), is for notational convenience taken to correspond to the time-domain equivalent of the frequency domain expansion given by
\begin{equation}
  \mathscr{F}[D(\omega)] = i\sum_{m \geq 2}\frac{\beta_m}{m!}\omega^m,
\end{equation}
where $\omega^n \to i^n\partial^n/\partial t^n$. Choosing now $K = -iD(\omega_0) = \beta(2\omega_0) - \beta_0 -\beta_1\omega_0$ where $\beta_0 = \beta(\omega_0)$ and $\beta(2\omega_0)$ are the propagation constants of the FF and the SH, respectively, and noting that $D(0) = 0$, we see that the evolution equations take a form where CW fields at neither the fundamental nor the second harmonic will experience any linear phase shifts during propagation. Moreover, since the phase mismatch $\Delta k  = 2\beta_0 - \beta(2\omega_0) = (\beta_0-\beta_1\omega_0) - K$ we may rewrite the single pass evolution equations for the $m^{th}$ roundtrip as
\begin{equation}
  \left[\partial_z - D_1 + \frac{\alpha_d}{2}\right]A_1 = i\kappa A_2A_1^*e^{-i\Delta k z}
  \label{eq:map1}
\end{equation}
\begin{equation}
  \left[\partial_z - D_2 + \frac{\alpha_d}{2}\right]A_2 = i\kappa A_1^2e^{i\Delta k z},
  \label{eq:map2}
\end{equation}
where $D_1 \equiv D(\omega)$ and $D_2 \equiv D(\omega_0) - D(\omega+\omega_0)$ are group-velocity dispersion operators for the fundamental and second-harmonic fields.

Using the ansatz (\ref{eq:ansatz}) in the boundary condition Eq. (\ref{eq:Ikeda1}) and projecting onto each frequency component similarly gives that
\begin{equation}
  A_1^{m+1}(t,0) = \sqrt{\theta}A_{in} + \sqrt{1-\theta}e^{-i\delta_1}A_1^m(t,L),
  \label{eq:map1b}
\end{equation}
\begin{equation}
  A_2^{m+1}(t,0) = \sqrt{1-\theta}e^{-i(\delta_1 - KL)}A_2^m(t,L),
  \label{eq:map2b}
\end{equation}
where the detuning at the second harmonic $\delta_2 = \delta_1 - KL = \delta_1 - (\beta(2\omega_0) - \beta_0 -\beta_1\omega_0)L$. Note that the detuning is generally a large quantity but that only the difference between the nearest integer of $2\pi$ is of significance. The detuning is in the absence of dispersion, i.e.~$K = 0$, equal to the pump detuning $\delta_1 = -\phi_0$ for all modes.

The map (\ref{eq:map1})-(\ref{eq:map2b}) models the evolution of a doubly resonant SHG cavity containing a uniform medium and can be directly compared with Ref.~\cite{Leo15b}. The quasi-phase-matching introduced by radially polling the resonator and the periodic flipping of the sign of the nonlinear coefficient can be accounted for by keeping the dominant terms in the Fourier expansion of the spatial variation of the nonlinear coefficient, which is equivalent to the substitution
\begin{equation}
  \kappa \to \frac{2}{\pi}\kappa e^{\pm i(2\pi/\Lambda) z},
\end{equation}
and gives the associated phase mismatch $\delta k = \Delta k - 2\pi/\Lambda$.

Note that the quadratic nonlinear coefficient $\kappa$ is given in a form where $\chi^{(2)}$ has dimensions of $[\textrm{m}/\sqrt{\textrm{W}}]$. To enable easier comparison with previous results we may rewrite the quadratic coefficient in units of $\textrm{W}^{-1/2}$ by including a conversion factor so that
\begin{equation}
  \kappa = \frac{\omega_0\chi^{(2)}}{2\tilde{n}_1c}\sqrt{\frac{2}{\tilde{n}_2c\epsilon_0}}
\end{equation}
where the second-order susceptibility now has dimensions of $[\textrm{m/V}]$ and we have defined the refractive indices $\tilde{n}_1 = n(\omega_0)$ and $\tilde{n}_2 = n(2\omega_0)$. To obtain agreement with Ref.~\cite{Leo15b} we should have $\chi^{(2)} \to 4\chi^{(2)}_{\textrm{eff}}$, with the squared modulus of the fields representing power rather than intensity.
We further remark that an expansion of the cubic polarization term gives a Kerr contribution at the fundamental frequency with the coefficient
\begin{equation}
  \frac{3\epsilon_0\chi^{(3)}}{4}\rho_0 = \frac{\omega_0 n_2}{c} = k_0n_2,
\end{equation}
which is the conventional expression for scalar plane wave propagation when $|A_1|^2$ is an intensity. Note that the transverse mode profile has been neglected in the derivation of the SEE model and that the Kerr coefficient therefore differs by a factor of $1/2$ from that of a waveguide such as, e.g., an optical fiber \cite{Okamoto_2006}.


\bibliography{frequency_combs}



\end{document}